# A Novel Compact and Selective Gas Sensing System Based on Microspherical Lasers

Behsan Behzadi, Ravinder K. Jain, Mani Hossein-Zadeh

*Center for High Technology Materials (CHTM), University of New Mexico, Albuquerque, NM*
*Corresponding author: mhz@unm.edu*

We propose the use of wavelength-dependent spectroscopic absorption in the evanescent field of ultra-compact whispering gallery mode microspherical lasers as selective gas sensors and estimate "few ppm" detection sensitivity using a simple model. We also propose a sensor configuration based on integrated multi-channel microspherical laser arrays capable of selective gas detection at low concentration, and demonstrate that a compact 14-channel device could generate a unique and distinguishable power distribution over channel outputs upon exposure to three different gases.

## I. Introduction

High sensitivity detection of trace gas species using compact sensors is a critical task for environmental monitoring, chemical weapon detection, breath analysis (for medical diagnostic) and industrial process monitoring [1,2]. While numerous compact opto-chemical and electro-chemical sensors have been developed during the past few decades for detection of variety of gases, there is still a demand for compact low-cost gas detectors and analyzers with faster response times and acceptable sensitivities. Several improvements -- such as absorption spectroscopy by multipass gas cells, cavity enhanced (resonant) spectroscopy, noise immune cavity enhanced optical heterodyne spectroscopy, and laser intra-cavity absorption spectroscopy – have been developed and implemented successfully in numerous gas detection and analysis systems [1]. Nevertheless, the cost, complexity, and the relatively large size of these systems have limited their deployment in field applications. In recent years developments of microresonators, microlasers, optical fibers and integrated waveguides functioning at wavelength ranges spanning the visible to the mid-IR spectral range has created new possibilities for the design and development of the new class of low-cost, compact and integrated gas sensors based on "molecular fingerprint-type" spectroscopic optical absorption. Developing compact and cost-effective optical platforms based on such novel compact microresonator devices and well-known optical detection techniques enables developing portable and affordable gas sensors for practical applications.

Laser intra-cavity absorption spectroscopy (LICAS) is a well-known technique for measuring the trace gas concentrations with high sensitivity [3]. This technique takes advantage of long interaction length of light with target molecules within the laser cavity and has been successfully tested in numerous experiments based on solid-state, semicomductor and fiber lasers [4]. The advent of high quality factor (high-Q) optical Whispering gallery mode (WGM) microresonators [5] has enabled scaling down the passive and active optical cavities to sub-mm size. The long photon life time combined with evanescent optical wave that resides outside a WGM microcavity results in a long interaction length (up to few meters) and therefore high sensitivity to its surroundings medium. As such these microcavities are excellent candidates for exploiting the sensitivity of resonant optical absorption sensing on a chip-scale. While using passive WGM microcavities for sensing requires precise locking of the desired resonant wavelength (defined by the absorption of the target molecule) to a narrow-linewidth laser, WGM microlasers naturally generate a narrow linewidth optical radiation when pumped by the proper pump wavelength (usually much shorter than the required lasing wavelength) [5]. As such not only the radiation is naturally emitted in the cavity modes, but also by choosing the proper cavity diameter, structural (host) material, active ion (dopant) and dopant density the laser wavelength can be selected near a specific spectral region for sensing. The recent demonstration microspherical (diameter <100 mm) WGM laser with a wavelength of 2.7 micron has shown that these small sources can be extended from visible and near-IR range (previously shown in variety of wavelengths) to mid-IR range that is well-known as molecular finger-print region [6,7,8].

Here using a simple model we analyze the power variations of a single mode WGM microspherical laser as a function of the absorption coefficient of its surround medium. Next we propose a system based on multi-wavelength microspherical laser arrays that can generate unique and distinguishable power distribution over given number of output channels upon exposure to different gases. Through an example we demonstrate that using microsphere lasers based on Er:ZBLAN glass active material a 14 channel system can easily detect and distinguish ammonia, acetylene and hydrogen sulfide at 4-40 ppm concentration in a gaseous solution.

## II. INTRACAVITY ABSORPTION IN MICROSPHERE LASERS

The WGM Microspherical laser (Fig. 1(a)) can be generally treated as a bidirectional ring laser (Fig. 1(b)). In a microspherical laser the circulating resonant power in the cavity (that is also the gain medium) interacts with molecules in the surrounding medium throughout the whole roundtrip (through evanescence field) while in a ring laser gain and loss are induced at different locations within the roundtrip inside the cavity. However the difference between distributed and localized interaction do not affect the overall relation between output power, pump power and the properties of the active cavity.

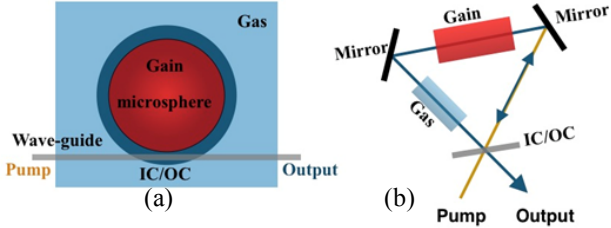

Fig. 1. (a) WGM microspherical laser (Diameter= 50-150 mm) surrounded by the target gas (light blue). The dark blue region shows the extent of the evanescent field (effective interaction region). (b) Ring laser with a gas cell.

Thus the theory of a "basic" bidirectional ring laser can be used for analyzing microsphere laser by redefining the roundtrip loss and gain. The output power can be simply written as [9]:

$$P_{out} = \frac{T P_{sat}}{2(1-S)} [\gamma_0 \mathcal{L} + \ln(S)] \quad (1)$$

In Eq. 1, T is the transmission coefficient of output coupler, $P_{sat}$ is the saturated power, $\gamma_0$ is the small signal gain, $\mathcal{L}$ is the length of the gain medium and S is the survival factor of the ring cavity which takes into account cavity losses and loss due to the gas absorption. The small signal gain is defined as:

$$\gamma_0 \mathcal{L} = A(P_{pump} - P_{min}) \quad (2)$$

where A is a constant that is function of pump absorption, efficiency and laser upper and lower level lifetimes. $P_{min}$ is the minimum required pump power to reach population inversion. In the case of a microspherical laser, transmission of the output coupler (T) is simply related to the roundtrip decay rate due to coupling between microsphere and waveguide ($k_c$) through $T = 1 - e^{-\kappa_c}$ The survival factor is related to the loss parameters and cavity properties through:

$$S \equiv 1 - \frac{1}{Q_{tot}} = e^{-\kappa_c - \alpha_s - \alpha_{gas}\Gamma\mathcal{L}} \quad (3)$$

where $\mathcal{L}$ is the circumference of the microsphere. $\alpha_{gas}$ is the absorption coefficient of the target gas defined by $a_{gas} = n\, s_{gas}$ where n is the gas concentration (cm$^{-3}$) and $\sigma_{gas}$ is gas absorption cross-section area. $\alpha_s$ is the round trip power reduction factor of the sphere due to material absorption, radiation loss and surface scattering. $\Gamma$ is the ratio of power circulating outside the sphere (evanescence wave interacting with the target gas environment) to the total intensity inside the sphere (gain medium) and is a function of sphere radius and laser wavelength. For a given WG mode $\Gamma$ can be calculated using numerical electromagnetic simulation. Fig. 2. shows the variation of the laser output power as a function of the absorption coefficient of the surrounding medium for a heavily doped Er:ZBLAN microsphere laser with a diameter of 80 micron, total quality factor of $10^7$ (in the absence of gas) emitting at 2.7 micron. The inset shows power variation near small concentrations.

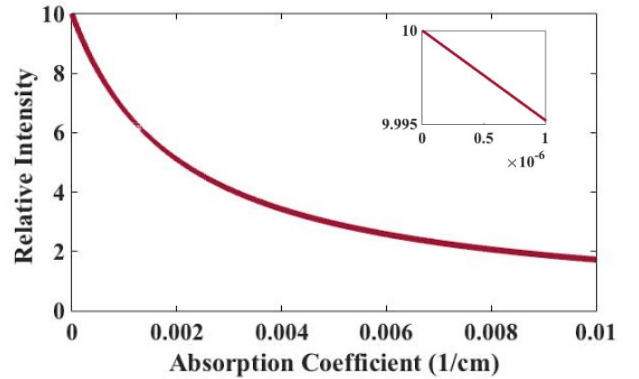

Fig. 2. Variation of the laser output power as a function of the absorption coefficient of the surrounding medium for a heavily doped Er:ZBLAN microsphere laser with a diameter of 80 micron, total quality factor of $10^7$ emitting at 2.7 micron. The inset shows power variation near small concentrations.

## III. Multi-wavelength Microspherical Laser Arrays For Selective Gas Sensing

A high-Q microsphere can be simply formed by melting a doped fiber through a low cost process using 1 cm fiber resulting in a total cost of less than 1 dollar per sphere. Once a microsphere is fabricated it can be side coupled to an integrated waveguide with the proper effective refractive index which couples the pump optical power (usually a near-IR DFB laser) to the microsphere and couples out the laser power (with a wavelength that depends on the dopant and its concentration). Integrated waveguides that are transparent from NIR to MIR range (the relevant range for gas sensing) have been already demonstrated [5,10]. Figure 3(a) shows the schematic diagram of proposed configuration where n microsphere lasers are coupled to n waveguides (channels). One (or two) lasers pump each microsphere and the laser outputs within NIR (MIR) range are combined and delivered to NIR (or MIR) photodetectors. To capture the molecule absorption near each available wavelength, multiple channels are dedicated to each wavelength (forming a group). The microsphere lasers within each group (shown with same color on the diagram) are made of the same active material but with slightly different diameter and pumped with the same wavelength. So each group will generate a slightly different wavelength within the gain bandwidth of the selected material (in Figure 3(a) each group has two microspheres with wavelength difference d).

Note that microspherical lasers can be fabricated based on the same host material and dopants used in fiber lasers. In principle all the wavelengths generated by fiber lasers can be also generated by microsphere lasers and therefore be used for intracavity gas sensing in a small form factor. So far a plethora of fiber lasers with wavelengths ranging from NIR (1.5, 1.94, 2.05, 2.1, 2.14, 2.3 micron) to MIR (2.7, 2.86, 2.9, 3, 3.22, 3.45, 3.95 micron) have been demonstrated [11]. While we expect that rapid development of new host materials (specially in MIR range) combined with variety of dopants will expand the available wavelengths, even with the existing wavelengths our proposed platform can be used for detection of several gases that are relevant for breath sensing [2] and environmental monitoring [1]. In the proposed configuration all the microspheres and waveguides are integrated in a small chip (20 channels can be fabricated on 1cm by 1cm chip). To obtain a baseline for the output power distribution over the channels in the absence of any target molecule the channels are sequentially pumped (by the proper pump wavelength) and the output power of each channel is measured by the two photodetectors. This will generate a "power-map" that simply shows the output power of each channel in the absence of molecules.

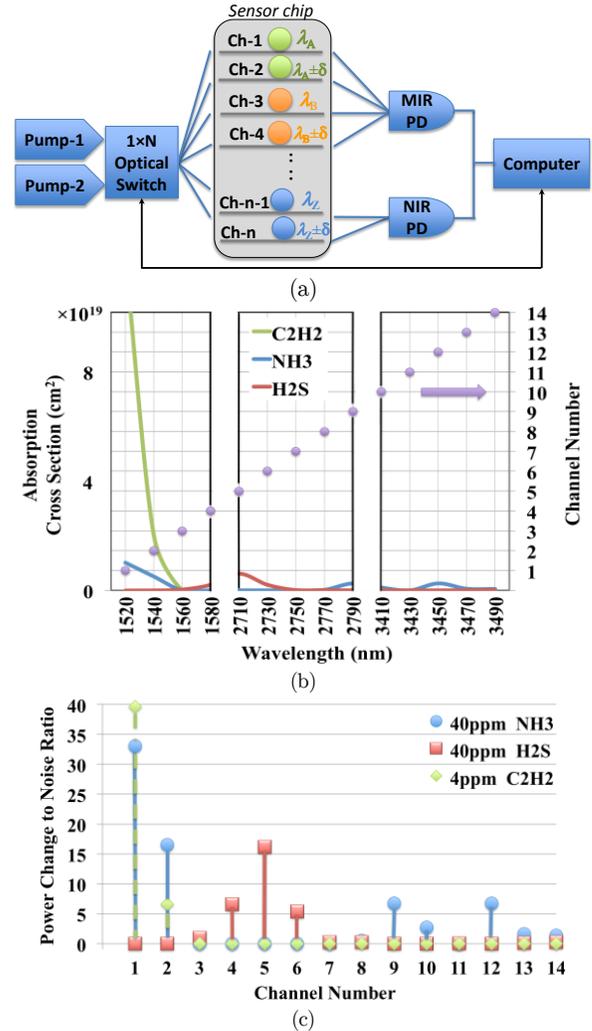

Fig. 3. (a) The proposed sensor configuration. In this case (b) The wavelength of microspheres coupled to each channel (purple dots) and the (c) Air pressure broadened absorption cross-section of three selected target gases (acetylene, hydrogen sulfide and ammonia) near each wavelength group [the cross-sections are extracted from HITRAN]. (c) Shows the estimated power-map for the system defined in part (b) exposed to a mixture of gases containing $C_2H_4$ at a concentration of 4 ppm, $H_2S$ at a concentration of 40 ppm and $NH_3$ at a concentration of 40 ppm. Here we assume that the signal-to-noise ratio of all the channels is 40 dB.

Next the sensor chip will be exposed to a known concentration of each target gas separately and the corresponding power-map will be recorded (by repeating the same sequential measurement process). Now if the sensor chip is exposed to an unknown mixture of gases the signature and possibly the concentration of each gas can be identified by comparing the measured power-map and the originally measured baseline and calibration power-maps. As an example we have selected $Er^{3+}$: ZBLAN glass that is capable of generating 1.5, 2.7 and 3.5 micron radiation depending on concentration of Er ions and pump wavelength (980 nm, 1900 nm or both). Moreover the microlaser can span the large portion of its gain medium only by changing its diameter. We assume four microsphere lasers with slightly different diameters are fabricated near each wavelength so that their wavelengths are 20 nm apart. So the resulting system will have 12 channels. Figure 3(b) shows the wavelength of microspheres coupled to each channels as well as the absorption cross-section of three selected target gases (acetylene, hydrogen sulfide and ammonia) near each wavelength group. Figure 3(c) shows the estimated power-map using the model developed in section II for the above mentioned system exposed to a mixture of gases containing $C_2H_4$ at a concentration of 4 ppm, $H_2S$ at a concentration of 40 ppm and $NH_3$ at a concentration of 40 ppm.

## IV. Conclusion

We have studied the application of microspherical lasers for evanescent-wave "intracavity" optical absorption sensing in a compact device. Using a simple model we have estimated the sensitivity of the optical output power of a microspherical laser upon exposure to a gas with absorption near the lasing wavelength. Next we have proposed a gas sensor based on arrays of microspherical lasers coupled to integrated waveguides that can be fabricated on a small "sensor chip". Using an example we have demonstrated upon exposure to a mixture of three different gases, this device can generate a power distribution that clearly reveals the type of the molecules present in the mixture. Note that implementation of the proposed configuration using any other laser would result in much more complex system. For example if one uses individual lasers the large number of expensive narrow-linewidth lasers (especially those generating MIR wavelengths) will make the system very expensive and large. Alternatively using tunable lasers and high resolution optical spectrometers (in particular when covering the MIR range) would make the system significantly larger and extremely expensive. Additionally tunable lasers cannot cover large wavelength ranges. The proposed system only requires one or two low-cost NIR DFB lasers (as pumps), two photodetectors and a sensor-chip (consisting of waveguides and lasers) that can be fabricated with relatively low cost.

## Acknowledgment

This work was supported by National Science Foundation (NSF Grant ECCS-1232263).